\documentclass[prb,twocolumn,aps,superscriptaddress]{revtex4-2}
\usepackage{graphicx}
\usepackage{dcolumn}
\usepackage{bm}
\usepackage{amssymb}
\usepackage{amsmath}
\usepackage{subfigure}
\usepackage{physics}
\usepackage{sublabel}
\usepackage[utf8]{inputenc}

\usepackage{hyperref}
\usepackage{floatrow}
\usepackage{graphicx}
\usepackage{bbm}
\hypersetup{colorlinks = true, linkcolor=blue, citecolor=blue, urlcolor=}
\usepackage[T1]{fontenc}

\begin{document}


\title{Conductance Spectroscopy of Majorana Zero Modes in Superconductor-Magnetic Insulator Nanowire Hybrid Systems
}

\author{Roshni Singh}
\affiliation{Department of Physics, Indian Institute of Technology Bombay, Powai, Mumbai-400076, India}

\author{Bhaskaran Muralidharan}
\affiliation{Department of Electrical Engineering, Indian Institute of Technology Bombay, Powai, Mumbai-400076, India}
\affiliation{Centre of Excellence in Quantum Information, Computation, Science and Technology, Indian Institute of Technology Bombay, Powai, Mumbai-400076, India}
\email{Corresponding author: Bhaskaran Muralidharan, \newline Email: bm@ee.iitb.ac.in}

\date{\today}
\begin{abstract}
There has been recent interest in superconductor-magnetic insulator hybrid Rashba nanowire setups for potentially hosting Majorana zero modes at smaller external Zeeman fields. Using the non-equilibrium Green's function technique, we develop a quantum transport model that accounts for the interplay between the quasiparticle dynamics in the superconductor-magnetic insulator bilayer structure and the transport processes through the Rashba nanowire. We provide an analysis of three-terminal setups to probe the local and non-local conductance in clean and disordered nanowires. We uncover the gap closing and reopening followed by the emergence of near-zero energy states, which can be attributed to topological zero modes in the clean limit. In the presence of a disordered potential, trivial Andreev bound states may form with signatures reminiscent of topological zero modes. Our results provide transport-based analysis of regimes that support the formation of Majorana modes in these hybrid systems while investigating the effect of disorder on devices.
 
\end{abstract}

\maketitle
\section{Introduction}
Rashba nanowire-superconductor hybrid systems \cite{Alicea-2010,Sau-2013,Stanescu_2013,PhysRevLett.105.077001,PhysRevLett.104.040502,Jiang_2013,Wilczek2012} are the front-running platforms for detecting and manipulating Majorana zero modes (MZMs) \cite{kitaev:physusp2001,Sarma2015,Aasen-2016,obrien:prl2018,RevModPhys.80.1083,prada2020andreev}. The quantized zero-bias conductance peak (ZBCP), observed in two terminal  setups featuring the normal metal - topological superconductor (N-TS) link, once considered to be a definitive signature of MZMs \cite{Das2012,Mourik-2012,Deng-2016,PhysRevLett.110.126406,Scaling_ZBP_Marcus,Albrecht2016,Scaling_ZBP_Marcus}, has become a controversial issue. Quasi-MZMs \cite{PhysRevB.100.045302,PhysRevLett.109.267002,PhysRevLett.109.227005,PhysRevB.96.201109,prada2020andreev,avila2019non}, which are near-zero energy trivial Andreev bound states (ABS) mimic most of the MZM signatures. \\
\indent As a result, recent efforts {\cite{10.21468/SciPostPhys.7.5.061,PhysRevB.86.100503,PhysRevB.97.155425,PhysRevB.96.075161,Lobos,PhysRevB.91.024514,San-Jose2016,awoga2019supercurrent}} have focused on  distinguishing between trivial and topological zero-energy modes. Recent efforts \cite{Akhmerov,PhysRevB.96.195418,PhysRevB.88.180507,Akhmerov,Flensberg_Nonlocal,Puglia_Cond_Matrix,puglia}, culminating in the topological gap protocol (TGP) \cite{pikulin2021protocol,aghaee2022inas}, have converged on guidelines that necessitate the measurement of all the elements of the conductance matrix, particularly focusing on non-local transport measurements using three-terminal normal-topological superconductor-normal (N-TS-N) setups to identify the bulk-gap closing and reopening, which separates the trivial and topological regimes. Non-local conductance signatures should thus supplement the local conductance measurements in identifying MZM signatures by detecting non-local correlations, particularly in the presence of disorder.\\

\begin{figure}
    \centering
    \includegraphics[scale = 0.06]{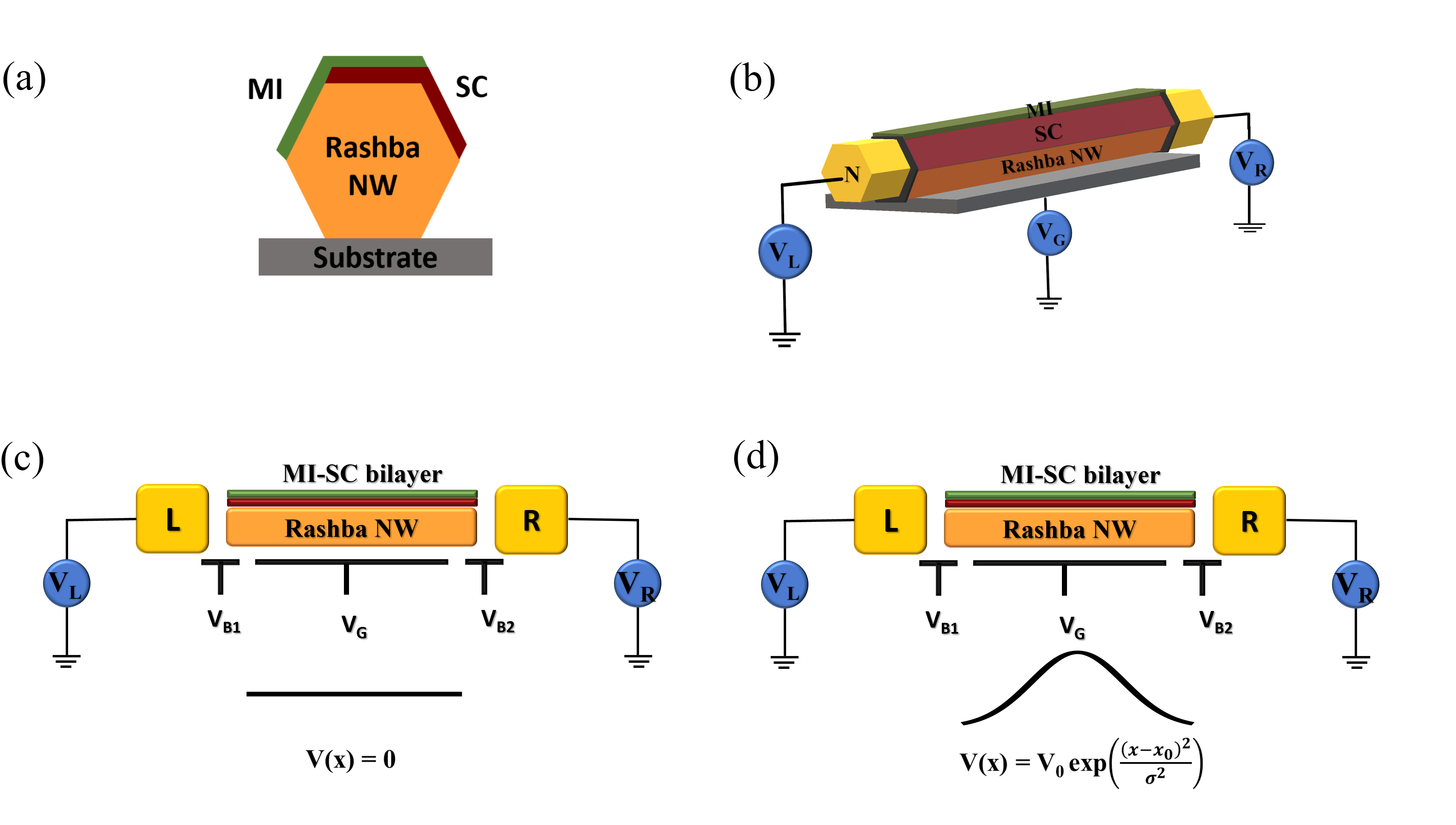}
           \caption{Device schematics. (a) Cross section of Rashba nanowire epitaxially epitaxially coated with a superconductor (SC) and a magnetic-insulator (MI), showing overlapping SC and MI layers. (b) A 3-D schematic of the device setup with the nanowire connected to two normal contacts via tunnel barriers, and a gate to control chemical potential, $\mu$.  (c) and (d) Effective 1D models used for computation, treating MI-SC as a stacked bilayer, with the homogeneous and inhomogeneous chemical potential profiles shown below.}
    \label{fig:device_setup}
\end{figure}
\indent With the aforementioned on one hand, the basic Rashba wire setup itself has further drawbacks which includes the requirement of large magnetic fields that could potentially destroy superconductivity \cite{ginzburg2009theory,sarma1963influence,chandrasekhar1962note,clogston1962upper} apart from the practicalities of precise magnetic field alignment \cite{karzig2017scalable}. Recently, efforts are being made towards realizing topological superconductivity with zero external magnetic fields by using proximity effects from magnetic insulators (MI) \cite{Yeyati_SCF,vaitiekenas2020zero,woods2020electrostatic,maiani2021topological,liu2021electronic,langbehn2021topological,khindanov2021topological,liu2019semiconductor,manna2020signature,escribano2021tunable,Yeyati_2022,zhang2020phase,vaitiekenas2022evidence,razmadze2022supercurrent}. Recent experimental \cite{vaitiekenas2020zero} and theoretical works \cite{woods2020electrostatic,maiani2021topological,liu2021electronic,langbehn2021topological,khindanov2021topological,escribano2021tunable,liu2022optimizing} featuring this setup indicate that at very low external magnetic fields, or even zero external magnetic fields, a topological phase can emerge. \\ 
\indent {Theoretical studies on the isolated system, schematized in Fig.~\ref{fig:device_setup}(a), thus far have set a preliminary stage by providing an insight on the geometric configurations and regimes of physical parameters which support the emergence of topological phases. However, the experimental basis for probing the formation of MZMs, in connection with the TGP, will entail a detailed analysis of conductance signatures. This necessitates detailed quantum transport calculations to evaluate the`current flow in a multi-terminal geometry that is schematized in Fig.~\ref{fig:device_setup}(b). } The object of this paper is hence to provide an in-depth analysis of the transport signatures of MZMs in these structures, particularly focusing on the local and non-local conductance spectra in both pristine and disordered nanowires.\\
\indent Using the Keldysh non-equilibrium Green's function (NEGF) technique, we develop a detailed quantum transport approach that accounts for the complex interplay between the quasiparticle dynamics in the superconductor-magnetic insulator (SC-MI) bilayer structure, and the transport processes through the semiconducting Rashba nanowire. Using this, we provide a detailed analysis of three terminal setups to probe the local and non-local conductance spectra in both the pristine as well as the disordered cases. We uncover the conductance quantization scaling with the bilayer coupling and the signatures of the gap closing followed by the emergence of near-zero energy states, which can be attributed to the zero modes in the clean nanowire. However, in the presence of a smoothly varying disorder potential, trivial Andreev bound states may form with signatures reminiscent of topological zero modes. 
\section{Results and Discussions}
\begin{figure}[t]
    \centering
    \includegraphics[scale = 0.12]{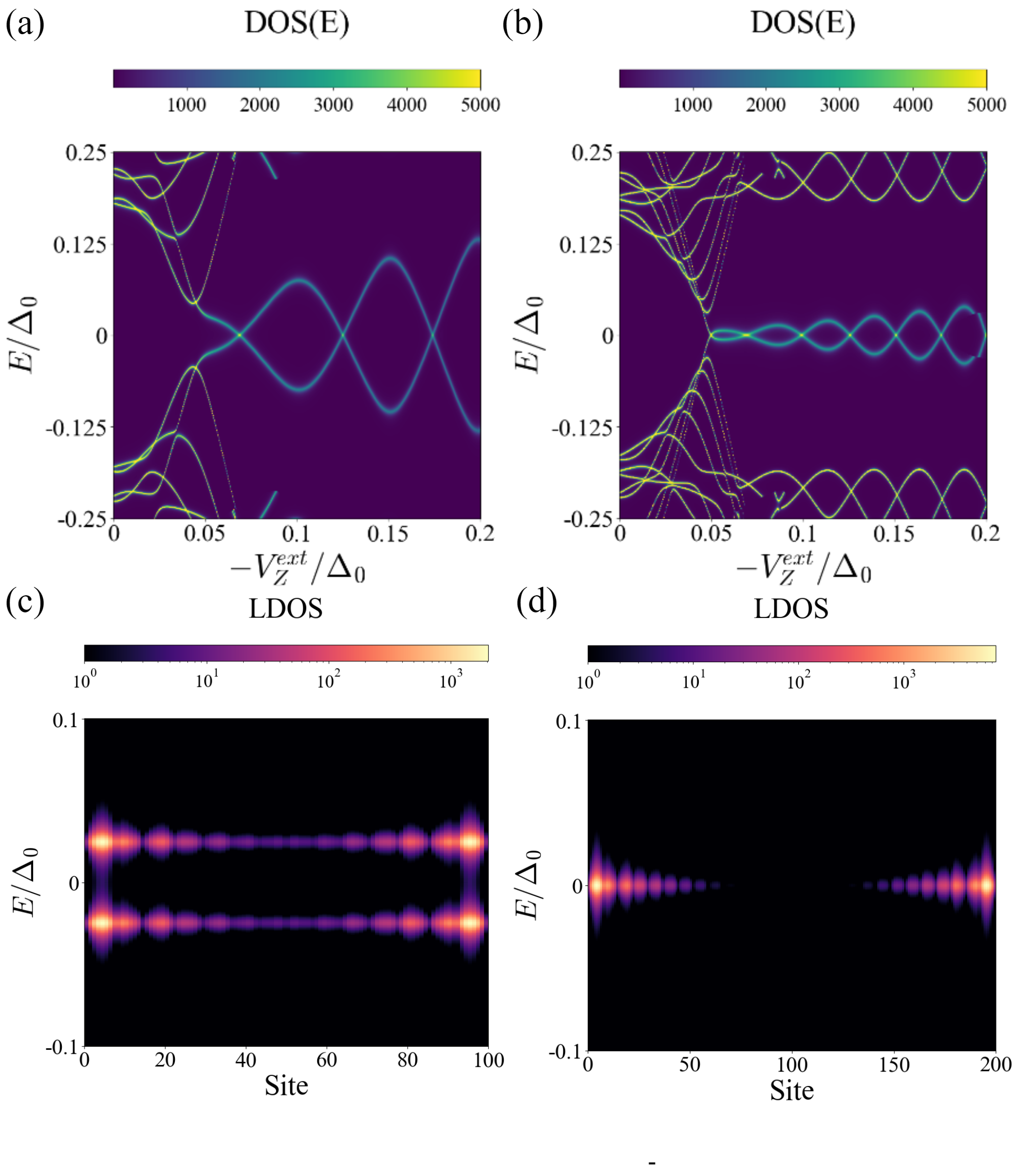}
          \caption{Majorana Zero Modes (MZMs) in the structure. (a), (b) Density of states (DOS) as a function of the energy(E) and the externally applied Zeeman field $V_{Z}^{ext}$ of the device region for nanowire (NW) lengths (a) $2.25 \mu m$, and  (b) $4.5 \mu m$. The low energy DOS shows the gap closing, followed by the emergence of a state near zero energy. The splitting of this low energy state is greater for shorter nanowires than longer nanowires. (c) and (d) Local density of states (LDOS) profiles at $V_Z^{ext} = 0.1\Delta_0$ which clearly show the localization of the zero energy states at the ends of the nanowire, consistent with the appearance of MZMs. The longer (length $4.5 \mu m)$ nanowire (d) shows a greater degree of localization than the shorter (length $2.25\mu m$) nanowire (c). The bare superconducting gap in the parent superconductor $\Delta_0$ sets the scale for all energies. The colorbars represent the magnitude of the DOS, LDOS.}
    \label{fig:Figure2}
\end{figure}
We consider semiconductor nanowires (SM) with Rashba-spin-orbit coupling with epitaxial layers of superconductors (SC) (usually Al/Pb) and magnetic insulators (MI) (usually EuS), as depicted in Fig.~\ref{fig:device_setup}(a). We then consider the device geometry where the MI and SC are in contact with the SM individually and overlap with each other, and connected to metallic leads as depicted in Fig.~\ref{fig:device_setup}(b). 

{The strength of the coupling of the metallic leads to the nanowire is controlled by the parameter $\gamma$, which represents the escape rate of the electrons into the leads. In the broadband limit of treating the contacts, as described in earlier works, this quantity is energy independent and can be used as a parameter. }The isolated Rashba nanowire is described by the following Hamiltonian:
\begin{equation}
    H_{SM} = V_Z^{SM}\hat{\sigma}_x + (\frac{\hbar^2k^2}{2m^*} - \mu + \alpha_R k \hat{\sigma}_y)\hat{\tau}_z,
\end{equation}
Where  $V_Z^{SM}$ is the Zeeman Hamiltonian in the SM, $\mu$ is the electrochemical potential, $\alpha_{R}$ is the strength of the Rashba spin-orbit coupling, $m^{*}$ is the effective mass of the electron and $\hat{\sigma}_{i},\hat{\tau}_{i}$, are the Pauli matrices in the spin and the particle-hole space, respectively. \\
\indent As detailed in the Methods section and in Supplementary Note 1, the effects of the SC-MI bilayer are accounted for as a self-energy term in the Green's function for the nanowire and the effect of the direct coupling of the MI to the nanowire is taken to be an effective Zeeman field in the wire. {The strength of the coupling of the SC-MI bilayer to the Rashba nanowire is controlled by the parameter $\gamma_{SC}$. It comes into play while calculating the self-energy accounting SC-MI bilayer from the Green's function for the bilayer. }We also use the self-consistent value of the superconducting gap, $\Delta$, calculated from the bare superconducting gap $\Delta_0$ of the parent superconductor in the presence of the Zeeman field and scattering processes. The process of obtaining this is involved and has been described in Supplementary Note 1. We use $\Delta_0 = 0.23$ meV, $m^{*} = 0.015 m_{e}$, where $m_{e}$ is the electron rest mass, for all our simulations. \\
\indent In order to model the system to simulate transport measurements, we reduce the hexagonal nanowire to a quasi one-dimensional system \cite{khindanov2021topological} as shown in Fig.~\ref{fig:device_setup} (c) and (d). {In Fig.~\ref{fig:device_setup}(c), we consider a `clean' nanowire, with a constant chemical potential. We also consider an inhomogeneous potential which is a spatially varying Gaussian potential as illustrated in Fig.~\ref{fig:device_setup}(d). Smoothly varying Gaussian potentials such as the one we have used in our simulations may arise in experimental situations due to charged impurities in the nanowire \cite{Pan-2020,penaranda2018quantifying}}.

\begin{figure*}[t]
    \centering
    \includegraphics[scale = 0.125]{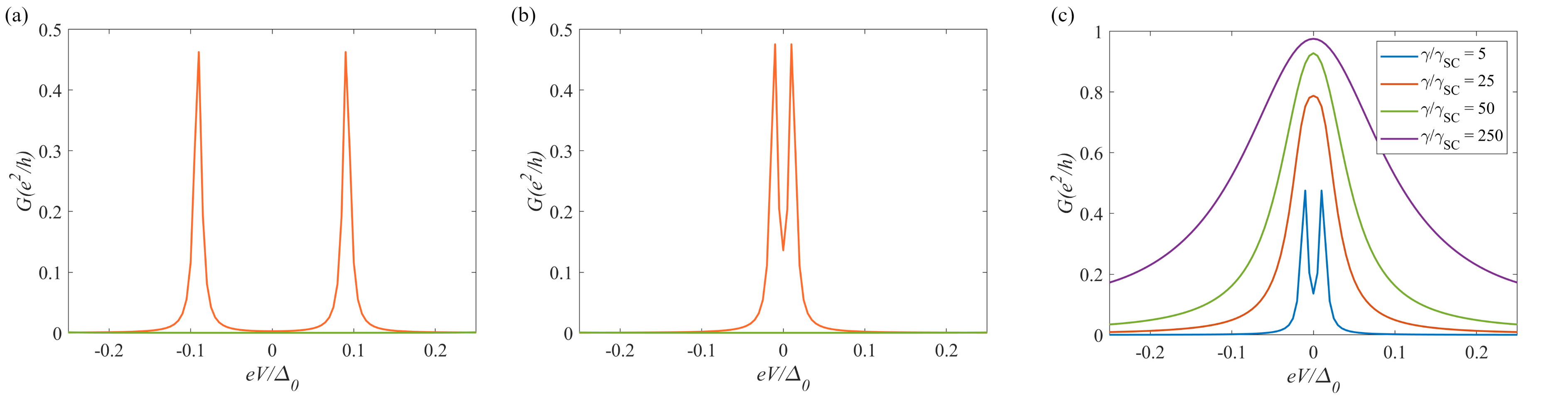}
    \caption{Scaling of conductance quantization with bilayer coupling. (a) and (b) Low bias differential conductance plots for nanowires of length (a) $2.25 \mu m$, and  (b) $4.5 \mu m$ in the topological region (External Zeeman field, $V_{Z}^{ext}=0.07\Delta_0$) (orange) and trivial region ($V_{Z}^{ext}=0.005\Delta_0$) (green). The topological regime shows clear conductance peaks absent in the trivial regime, though not quantized. The splitting of the zero bias peak is more pronounced for the shorter nanowire, consistent with Fig.~\ref{fig:Figure2}.(c) Shows that as the coupling to the normal contacts, $\gamma$, is increased, so that it becomes much larger than the coupling, $\gamma_{SC}$, between the nanowire and the superconductor-magnetic insulator bilayer, the peak asymptotically reaches the expected quantized value. The bare superconducting gap in the parent superconductor $\Delta_0$ sets the scale for all energies. The differential conductance, $G$ is plotted as a function of the bias, $V$. G is measured in units of $e^2/h$, which is the conductance quantum, $e$ being the electronic charge and $h$ being Planck's constant.}
    \label{fig:Figure3}
\end{figure*}
\indent An external Zeeman field is applied which is anti-parallel to the magnetization in the MI. It reduces the Zeeman term in the Hamiltonian of the SC, but increases the Zeeman field in the normal metal. We parameterize the Zeeman fields in the SC and the SM in terms of the field directly induced in the SC due to the MI $(V_0^{SC})$, and the coupling strengths of the SC and SM nanowire to the external magnetic field, $(g_{SC},g_{SM})$ as follows: 
\begin{equation}
    \begin{split}
        V^{SC}_Z &= V^{SC}_0 + g_{SC}V^Z_{ext}\\
        V^{SM}_Z &= g_{SM}V^Z_{ext}.
    \end{split}
    \label{eq:Ham}
\end{equation}

We use $g_{SC} = 2, g_{SM} = -15$ for our calculations, closely following the setup in \cite{khindanov2021topological}, used for equilibrium calculations. {A  finite Zeeman energy in the semiconductor, either induced by coupling to a magnetic insulator or by an applied magnetic field, is required to enter the topological phase \cite{poyhonen2021minimal}. Here we have considered only an external Zeeman field, and neglected the direct coupling of the SM to the MI - an assumption which can altered by adjusting the parameterization.} \\

\begin{figure}[t]
    \centering
    \includegraphics[scale = 0.16]{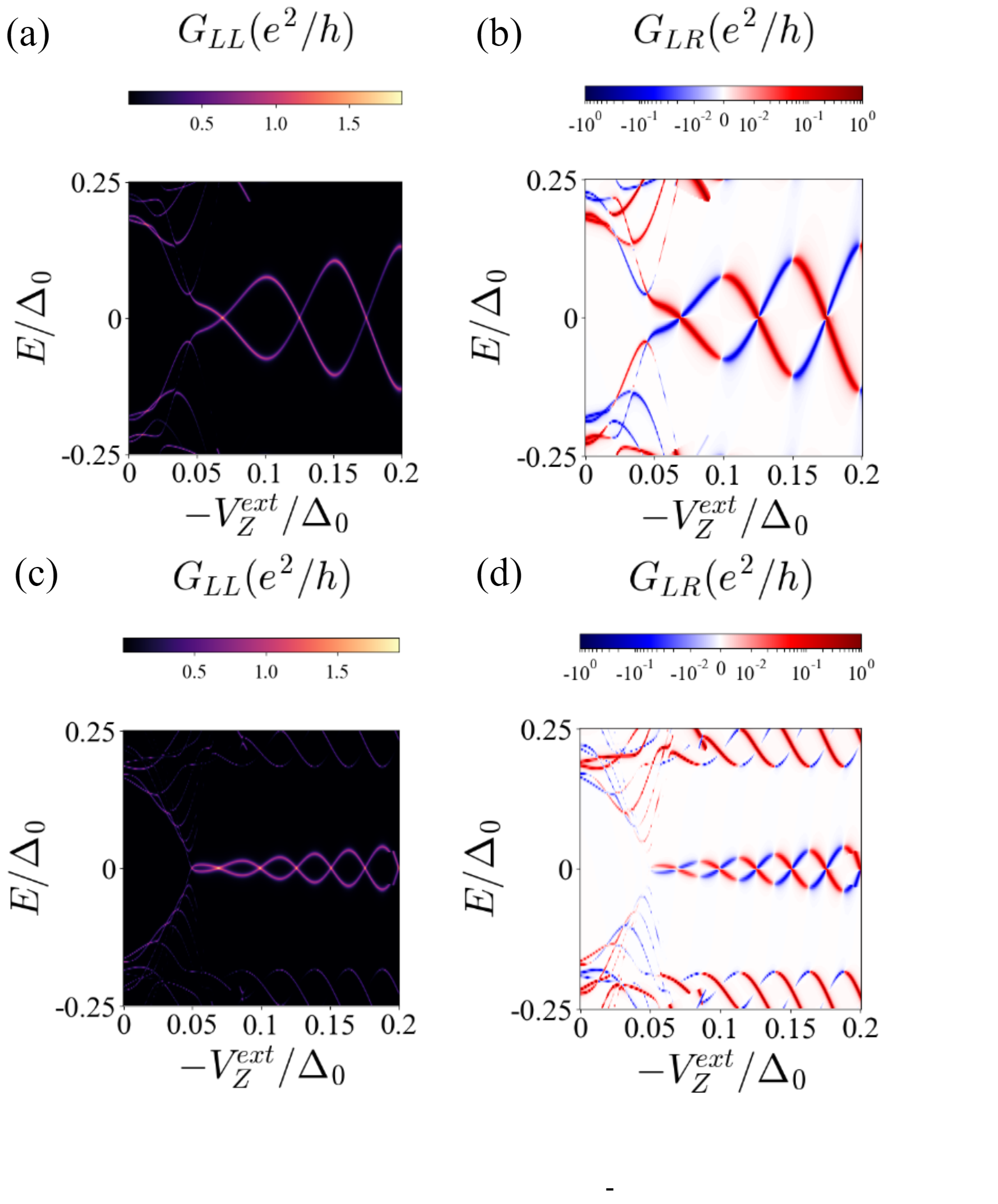}
        \caption{Conductance spectra for the clean nanowire. (a) Local ($G_{LL}$) and (b) non-local conductance ($G_{LR}$) signatures for a nanowire of length 2.25 $\mu m$, with a potential profile as shown in Fig. ~\ref{fig:device_setup}(c). (c) Local and (d) non-local conductance spectra for a nanowire of length 4.5 $\mu m$, with a potential profile as shown in Fig.~\ref{fig:device_setup}(c). Both the local and the non-local conductances show the bulk gap closing and reopening, which signals a topological transition, followed by the emergence of a near-zero energy state with a splitting that oscillates as a function of the externally applied magnetic field, $V_{Z}^{ext}$. The bare superconducting gap in the parent superconductor $\Delta_0$ sets the scale for all energies. $G_{LL}, G_{LR}$ are plotted as a function of the biasing energy (E), and the externally applied magnetic field $V_{Z}^{ext}$ and are measured in units of $e^2/h$, which is the conductance quantum, $e$ being the electronic charge and $h$ being Planck's constant. The colorbars represent the magnitude of the DOS, LDOS.}
    \label{fig:Figure4}
\end{figure}
\begin{figure*}[!t]
    \centering
    \includegraphics[scale = 0.18]{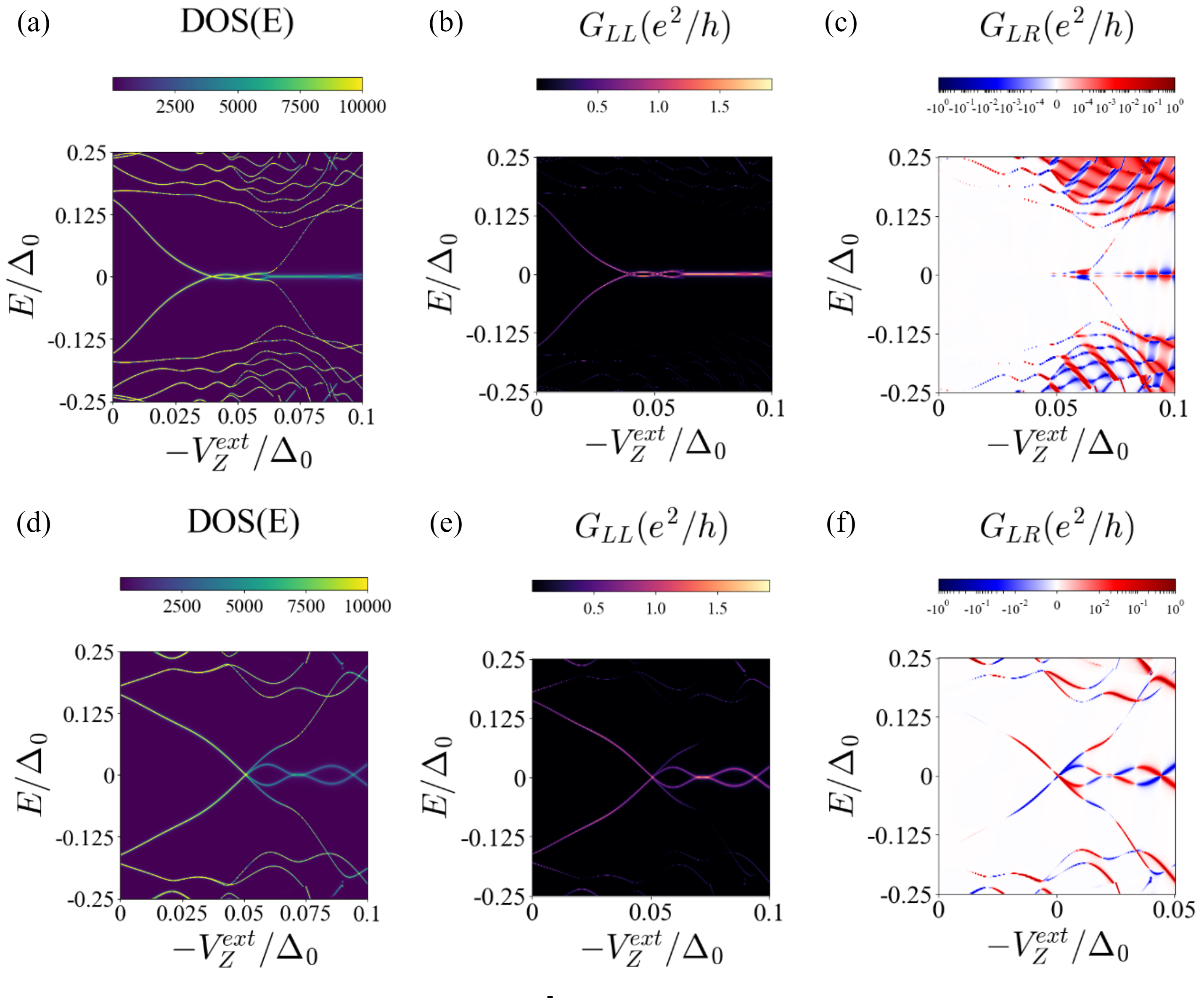}
    \caption{Conductance spectra for the disordered nanowire. (a) Low Energy density of states (DOS) (b) local and (c) non-local conductance spectra for a nanowire of length 4.5 $\mu m$, with a potential profile as shown in Fig.~\ref{fig:device_setup}(d). (d) Low energy DOS, (e) the local and (f) the non-local conductance conductance spectra for a nanowire of length 2.25 $\mu m$, with a potential profile as shown in Fig.~\ref{fig:device_setup}(d). For the long nanowire, we see premature emergence of a quasi-Majorana Zero Mode before the bulk gap closing and reopening.The bare superconducting gap in the parent superconductor $\Delta_0$ sets the scale for all energies. $G_{LL}, G_{LR}$ are plotted as a function of the biasing energy (E), and the externally applied magnetic field $V_{Z}^{ext}$ and are measured in units of $e^2/h$, which is the conductance quantum, $e$ being the electronic charge and $h$ being Planck's constant. The colorbars represent the magnitude of the DOS, LDOS.}
    \label{fig:Figure5}
\end{figure*}
\indent We first present the numerical results for the pristine nanowire, in the absence of an inhomogeneous potential. The low energy density of states (DOS) shown in Figs.~\ref{fig:Figure2} (a) and (b) illustrates the lowest ABS which form a near-zero energy oscillating mode after the topological transition, which is marked by the bulk gap closing and reopening. The bulk gap closing and reopening is more prominent in the longer nanowire than the shorter one, since there are more sub-gap states. As the length of the nanowire increases, the oscillations around zero-energy are exponentially suppressed\cite{Duse_2021,leumer2020linear,kejriwal2021can}. \\
\indent The local density of states (LDOS) corresponding to these nanowires, as seen in Figs.~\ref{fig:Figure2} (c) and (d), shows that the zero energy state is well localised at both ends, and more prominently so in the longer nanowire. The LDOS plots also show a greater splitting in energy around the zero energy for the shorter nanowire than the longer nanowire. While on closer inspection, the longer nanowire also shows some splitting, in reasonable experimental measurements we expect to see a single broadened peak. These observations are due to the hybridization of the MZMs when they overlap in finite nanowires, resulting in a splitting of the zero mode \cite{sarma2012splitting}. The hybridization of the MZMs through the nanowire is suppressed with increasing length \cite{Duse_2021,leumer2020linear,kejriwal2021can}, which is consistent with our observations. \\
\indent In Figs.~\ref{fig:Figure3}, (a) and (b) we plot the differential conductance for clean nanowires of lengths $2.25 \mu m$ and $4.5 \mu m$ with chemical potential $\mu = 0.125 $meV and find a clear ZBCP in the topological regime, and no peak near zero energy in the topologically trivial regime. The peak is close to the quantized value of $\frac{e^2}{h}$ expected from an MZM in an N-S-N setup under symmetric biasing \cite{leumer2020linear,Duse_2021}, but is smaller. {We attribute this observation to the level broadening due to the composite SC-MI bilayer which effectively acts as an extra contact \cite{datta2005quantum} and induces further broadening compared to a two-terminal N-TS-N device \cite{flensberg2021engineered}}. This is also borne out by the fact that as the coupling to the metallic contacts becomes much stronger than the coupling to the bilayer, the ZBCP asymptotically reaches the quantized value, as shown in Fig.~\ref{fig:Figure3} (c), since the broadening due to the effective MI-SC bilayer becomes negligible in comparison to the broadening induced by the metallic contacts. It should also be noted that the exactness of the quantization would also depend on the external magnetic field because the Majorana overlap energy oscillates with the externally applied magnetic field. As the overlap energy of the MZMs at the ends of the nanowire varies, the peak value of the ZBCP also changes \cite{lai2021quality}. This figure clearly elucidates the effect of introducing the bilayer on the actual conductance quantization of MZMs in the setup. \\
\indent The ZBCP quantization reflects upon the local Andreev reflection process, and can be a diagnostic for MZMs in pristine setups. In disordered setups, as currently encountered in experiments, the current guidelines on the detection of MZMs follows the TGP \cite{pikulin2021protocol,aghaee2022inas}. The TGP employs a combination of local and non-local conductance measurements on three terminal setups as an effective diagnostic tool for the detection of MZMs. In this protocol, first, the local conductance measurements at the two ends of a disordered setup must individually show a sustained gap closure and ZBCP spanning a large range of applied magnetic fields. Such gap closures will have local variations due to the presence of random uncorrelated impurities at the two ends. In addition, the non-local conductance gap must close and re-open, signalling a topological phase transition. Besides that, at the point at which the non-local conductance gap closes, one must observe a zero magnitude of the conductance along with a flip in the sign of the conductance. From the transport physics angle \cite{Akhmerov,kejriwal2021can}, this can be attributed to the fact that the non-local conductance magnitude is given by the difference between the direct component and the crossed Andreev component, all of which are described in Supplementary Note 2 and several other references. \\
\indent We now turn our attention to our simulations on the local and non-local conductances in accordance with the TGP. Following previous analysis \cite{Pan-2019-prb,Pan-2020} for the disordered setup, we have specifically considered the inhomogenous potential as depicted in Fig.~\ref{fig:device_setup}(d). A smoothly-varying potential barrier can be interpreted as a spatially varying chemical potential. In such a scenario, parts of the nanowire may locally enter the topological regime, resulting in a local pair of near-zero energy modes.  This typically happens when the system is tuned close to the topological phase. These quasi-MZMs may mimic many of the characteristic signatures of MZMs, including zero bias peaks in the local conductance at magnetic fields smaller than those at which the topological MZMs appear. However, these near-zero-energy states do not have topological protection and are hence not true MZMs. \\
\indent {Before embarking on the results related to the local and nonlocal transport spectroscopy simulations, we must make a quick remark on the biasing conditions of three terminal setups. In such a setup, voltages can be applied and currents can be measured across either terminals independently. As defined in \eqref{eq3} and explicitly derived in \eqref{4} and \eqref{eq5}, this tantamounts to enforcing asymmetric biasing at the contacts, $(V_{L} = V, V_{R}=0)$. As a consequence, the ZBCP is quantized at $\frac{2e^{2}}{h}$, unlike that in Fig.~\ref{fig:Figure3}, where we had considered a symmetric bias scenario ($V_{L} = V/2,V_{R} = -V/2$), and hence obtained the ZBCP quantized at $\frac{e^{2}}{h}$ \cite{leumer2020linear}. The quantized peak is a signature of the MZM and may be attributed to a perfect and coherent Andreev reflection at a semiconductor - topological superconductor (N-TS) interface, typically seen in two-terminal devices.\cite{flensberg2021engineered}\\}
\indent Both the local and non-local conductance spectra for the pristine nanowire are shown in Fig.~\ref{fig:Figure4}, for nanowires of length $2.25 \mu m$ (Figs.~\ref{fig:Figure4}(a),(b)) and $4.5 \mu m$ (Figs.~\ref{fig:Figure4}(c),(d)) which exhibit similar features, that is, the bulk gap closing and reopening followed by the emergence of Majorana oscillations around zero energy. Since the zero modes appear after the closure of the bulk gap, we can conclude that they are not quasi-MZMs, but may indeed be topological MZMs \cite{puglia2021closing}. We also note that a finite low-bias non-local conductance only emerges after the topological transition. The low bias non-local conductance is rectifying in nature \cite{Akhmerov} and switches sign as the the voltage polarity is reversed. At the turning points, the non-local conductance vanishes.\\
\indent In the sub-gap region, there is a correspondence between the non-local conductance and the BCS charges of the bound state at the leads { such that the nonlocal conductance is proportional to the BCS charge at a contact  \cite{danon2020nonlocal}. The charge difference between the even and odd parity ground states is equal to the net charge carried by the BdG fermion state, which can be nonzero when the constituent Majorana wave functions overlap \cite{dominguez2017zero}. }The vanishing of the non-local conductance at the turning points is a signature expected from hybridized MZMs, which should be chargeless at turning points. {From the plots of local and nonlocal conductance of the pristine nanowire, we can see an oscillation in the energy splitting as the magnetic field is increased. Meanwhile, the magnitude of the nonlocal conductance, which corresponds to the BCS charge, vanishes at the turning points (maxima and minima) of the energy splitting, and is maximum (close to the quantized value) at the points where the energy splitting vanishes \cite{hansen2018probing}. Therefore, the BCS charge and the energy splitting oscillate out of phase with each other, which is a characteristic signature of MZMs. } The local conductance is almost quantized at $\frac{2e^2}{h}$ where the Majorana splitting goes to zero. The deviation from the precise quantization value may be attributed to contact broadening, and due to the fact that we have three contacts in our system, as elaborated previously. At the points where the Majorana overlap energy becomes significant, the value of the local conductance drops further.\\
\indent In presence of a smoothly-varying potential barrier, the formation of quasi-MZMs is expected \cite{hess2021local,kells2012near,prada2012transport,liu2017andreev,Vulik_2016,rosdahl2018andreev,reeg2018zero,penaranda2018quantifying}. They characteristically appear in the topologically trivial regime before the closure of the bulk gap, mimicking many signatures originally considered as `smoking gun signatures' for MZMs including oscillations with the externally applied magnetic field, and with the associated local conductance quantized at values close to $\frac{2e^2}{h}$.  It is expected that systems which show quasi-MZMs, will also exhibit true MZMs in the topological phase on increasing the external Zeeman field. For such a disordered case, for the longer nanowire, as shown in Fig.~\ref{fig:Figure5}(a), in the DOS, we find signatures characteristic of a quasi-MZM state, followed by a gap reopening signature and the emergence of a potential topological MZM. The local conductance, as shown in Fig.~\ref{fig:Figure5}(b) in this case is quite deceptive since we see a premature gap closing and the bulk-gap reopening signature is extremely faint. \\
\indent The quasi-MZM and the true MZM regions are quite difficult to distinguish. In the non-local conductance plot shown in Fig.~\ref{fig:Figure5}(c), the bulk-gap reopening is seen more prominently, in closer accord with the gap protocol \cite{pikulin2021protocol}. The non-local conductance shows signatures of both the quasi-MZM and the topological MZM states, which can be distinguished by their position with respect to the reopening of the bulk gap  \cite{hess2021local}. For the shorter nanowire, as seen in Fig.~\ref{fig:Figure5}(d),(e) and (f), the local and the non-local conductance spectra both show the gap closing and reopening followed by the emergence of MZMs, which oscillate in energy as the Zeeman field is increased. At the points where the splitting is zero, the local conductance is quantized at values very close to $\frac{2e^2}{h}$. We do not find any signatures of quasi-MZM states in this device. An interesting point to note is that the local conductance exhibits signs of negative differential conductance. It is also worth noting that especially for longer nanowires, neither the local nor the non-local conductance alone has the entire information regarding the channel DOS, that arises from its eigenspectrum.\\
\indent The MZMs are protected by a clear topological gap both for the pristine nanowire and for the nanowire with a smoothly varying background potential.  The local conductance fails to probe the bulk states for sufficiently long nanowires. Before one lays claim to having observed topological MZMs, it is necessary to measure the entire conductance matrix to probe a device and investigate whether the zero bias peaks at both the contacts or on both the sides are correlated and emerging after the bulk gap closing and reopening, \\
\indent To conclude, using the NEGF technique, we developed a detailed quantum transport approach that accounts for the complex interplay between the quasiparticle dynamics in the SC-MI bilayer structure, and the transport processes through the semiconducting Rashba nanowire. We provided a detailed analysis of three terminal setups to probe the local and non-local conductance spectra in both the pristine as well as the disordered limits. We uncovered the conductance quantization scaling with the bilayer coupling and the signatures of the gap closing followed by the emergence of near-zero energy states, which can be attributed to the topological zero modes in the clean nanowire limit. However, in the presence of a smoothly varying potential, trivial Andreev bound states may form with signatures reminiscent of topological zero modes in the form of a premature gap closure in the conductance spectra. Our results therefore provide transport-based analysis of the operating regimes that support the formation of MZMs in these hybrid systems of current interest, while considering experimentally relevant device structures with realistic disordered potentials accounting for shallow tunnel barriers which may be formed inside the hybrid nanowire structure. Having set the stage for understanding device modeling in these emerging structures, the technique can also be easily extended to account for other experimental device designs and also the inclusion of scattering effects \cite{doi:10.1063/1.5044254,Duse_2021} inside the nanowire channel. 
\section{Methods}
We discretize the Hamiltonian of the system \eqref{eq:Ham} on a 1D lattice with N sites, and write the Green's function in the Nambu spinor basis \cite{Praveen} $(\psi_{\uparrow},\psi_{\downarrow},-\psi_{\uparrow}^{\dagger},\psi_{\downarrow}^{\dagger})^T$. 
The Hamiltonian for the Rashba nanowire and the self-energies corresponding to the metallic contacts and the SC-MI bilayer, are then used obtain the retarded Green's function for the hybrid device which is used for our transport calculations,
\begin{equation} G^R =\left[ (E + i\eta)\mathbb{I} - H_{SM}  - \Sigma_L - \Sigma_R - \Sigma^\prime \right]^{-1}, \end{equation}
where $\eta$ is an infinitesimal positive damping parameter introduced for numerical stability, and $\mathbb{I}$ is the identity matrix of the dimension of the Hamiltonian matrix in Nambu space. In the wide-band approximation \cite{Duse_2021,leumer2020linear,kejriwal2021can}, the self energies for the metallic contacts, $\Sigma_{L(R)}$, are written in their eigenbasis and are hence diagonal, as detailed in Supplementary Note 2. 
We use the Usadel equation, which is derived from a quasi-classical approximation to the Gorkov equations, to find the Green's function, and hence, the self-energy, $\Sigma^{'}$ for the SC-MI bilayer \cite{khindanov2021topological}. The effect of the proximity of the MI on the SC can be taken into account in the boundary conditions of the Usadel equation. The MI layer induces a uniform Zeeman field $V_Z^{SC}$ in the diffusive superconductor\cite{cottet2009spin}. The Usadel equation is then solved with the self-consistent value of $\Delta$ to obtain the quasi-classical Green's function for the bilayer, $\check{g}$. The gap is induced in the bare Rashba nanowire by considering the proximity effect of the bilayer, which is taken into account using the self energy, $\Sigma^{'}$ which can be obtained from the semi-classical Green's function, $\check{g}$. The imaginary part of $\Sigma^{'}$ connects the electron and hole subspaces, thus, inducing a gap in the system.\\
\indent We also take into account spin-orbit and spin-flip scattering in the SC, by adding a scattering self-energy term in the Usadel equation. The energy scales for the spin-orbit and spin-flip relaxation processes are characterised by $\Gamma_{so}, \Gamma_{sf}$ respectively. We take $\Gamma_{so} = \Gamma_{sf} = 0.4\Delta_0$ for our simulations unless stated otherwise. We also use the Usadel Equation to calculate the self-consistent value of the superconducting gap in the presence of an external magnetic field. For this, we solve the Usadel equation self-consistently with the superconducting gap equation along with a thermodynamic constraint as outlined in Supplementary Note 1.
\\
\indent {The retarded Green's function is be used to calculate the spectral function, A(E), the trace of which gives the density of states (times $2\pi$), and the diagonal elements of which gives us the local density of states (times $2\pi$).}
\begin{equation}
    DOS(E) = \frac{1}{2\pi}Tr [A(E)] = \frac{1}{2\pi}Tr[G^{R}(E) - G^{A}(E)]
\end{equation}
\\
\indent We also use the retarded Green's function defined above to calculate the {transmission coefficients, and hence the} conductance matrix for this setup \cite{leumer2020linear,Duse_2021,kejriwal2021can,Praveen}. As shown in Fig.~\ref{fig:device_setup}, we apply voltages $V_{L(R)}$ to the left and right contacts and measure terminal currents $I_{L(R)}$. We use the Keldysh non-equilibrium Green's function formalism to evaluate the terminal currents \cite{leumer2020linear,Duse_2021,kejriwal2021can,Praveen}. The terminal electronic current at the left contact \cite{Datta} can be derived  in the Landauer B\"{u}ttiker form as:

\begin{equation} \label{current}
\begin{aligned}
I_{L}^{(e)}=&-\frac{e}{h} \left\{\int d E T_{A}^{(e)}(E)\left[f\left(E-e V_{L}\right)-f\left(E+e V_{L}\right)\right]\right.\\
&+\int d E T_{C A R}^{(e)}(E)\left[f\left(E-e V_{L}\right)-f\left(E+e V_{R}\right)\right] \\
&\left.+\int d E T_{D}^{(e)}(E)\left[f\left(E-e V_{L}\right)-f\left(E-e V_{R}\right)\right]\right\} + I',
\end{aligned}
\end{equation}
where, $T^{(e)}_{D}(E)$, $T^{(e)}_{A}(E)$, and $T^{(e)}_{CAR}(E)$ represent the energy resolved transmission probabilities for the direct, Andreev and crossed-Andreev processes involving the left and right contacts for the electronic sector of the Nambu space and I' is the extra current due to the SC-MI bilayer acting as an effective contact, derived in Supplementary Note 1.{ These transmission probabilities can be calculated from Green's function of the device and the self-energies of the contacts \cite{kejriwal2021can,Duse_2021}}.\\
\indent Using the expressions for the terminal currents from above, the conductance matrix $[G]$ can be defined as:
\begin{equation} \label{eq3}
\mathrm{[G]}=\left(\begin{array}{cc}
G_{L L} & G_{L R} \\
G_{R L} & G_{R R}
\end{array}\right)=\left(\begin{array}{ll}
\left.\frac{\partial I_{L}}{\partial V_{L}}\right|_{V_{R}=0} & \left.\frac{\partial I_{L}}{\partial V_{R}}\right|_{V_{L}=0} \\
\left.\frac{\partial I_{R}}{\partial V_{L}}\right|_{V_{R}=0} & \left.\frac{\partial I_{R}}{\partial V_{R}}\right|_{V_{L}=0}
\end{array}\right),
\end{equation}
The diagonal matrix elements represent the local conductance at the left and right contacts, and the off-diagonal components represent the non-local conductance. \\
\indent
The local conductance at the left contact can be derived by taking a partial derivative of the left terminal current ($I_L$), as given in \eqref{current}, with the left contact voltage ($V_L$), and the right contact voltage ($V_R$) set to zero, and is given by : $G_{LL}=\left.\frac{\partial I_{L}}{\partial V_{L}}\right|_{V_{R}=0}$. Using this, we derive the following expression for the local conductance using the Landauer B\"{u}ttiker form

\begin{equation} \label{4}
\begin{aligned}
\left.G_{LL}(V)\right|_{T \rightarrow 0} \equiv \frac{e^2}{h}\left[T_{A}(E=e V)+T_{A}(E=-e V)\right.+\\
\left.T_{C A R}(E=eV)+T_{D}(E=e V)+G_{LL}^{'}(V)\right],
\end{aligned}
\end{equation}

The term $G_{LL}^{'}(V)$ is due to currents flowing into the SC-MI bilayer\cite{kejriwal2021can}. \\
\indent
The non-local conductance formula can similarly be derived by taking a partial derivative of the left terminal current ($I_L$), as given in \eqref{current}, over the right terminal voltage ($V_R$), with the left terminal voltage ($V_L$) set to zero, such that,  $G_{LR}=\left.\frac{\partial I_{L}}{\partial V_{R}}\right|_{V_{L}=0}$.

\begin{equation} \label{eq5}
\begin{aligned} 
\left.G_{LR}(V)\right|_{T \rightarrow 0} \equiv \frac{e^2}{h}\left[T_{CAR}(E=-e V)-T_{D}(E=e V)\right].
\end{aligned}
\end{equation}
Using the above, we have analysed the local and non-local conductances of the device in both the pristine and disordered setups.

\section*{Data Availability}
The data that support the plots within this paper and other findings of this study are available from the corresponding author upon reasonable request.
\section*{Code Availability}
The codes generated during the simulation study are available from the corresponding author upon reasonable request.
\section*{Acknowledgements}
The author BM acknowledges the Visvesvaraya Ph.D Scheme of the Ministry of Electronics and Information Technology (MEITY), Government of India, implemented by Digital India Corporation (formerly Media Lab Asia). The author BM also acknowledges the support by the Science and Engineering Research Board (SERB), Government of India, Grant No. STR/2019/000030, and the Ministry of Human Resource Development (MHRD), Government of India, Grant No. STARS/APR2019/NS/226/FS under the STARS scheme. 
\section*{Author Contributions}
BM and RS conceived the idea of this project. RS performed all numerical simulations. Both the authors contributed in analyzing the results and writing the paper.
\section*{Competing Interests}
The authors declare that there are no competing interests.
\bibliography{main.bib} 
\newpage
\onecolumngrid
\section*{Supplementary Material}
\renewcommand{\theequation}{S\arabic{equation}}
\renewcommand{\thefigure}{S\arabic{figure}}
\setcounter{equation}{0}
\setcounter{figure}{0}
\onecolumngrid

The Usadel equation is the standard way to describe superconductors in the diffusive limit, yielding a quasiclassical Green's function which we use to compute the self-energy of the SC-MI bilayer in our setup. It is valid when the mean free path is less than the superconducting coherence length but still much greater than the metallic Fermi velocity. This approximation is reasonable for superconductors such as Al, which are commonly used in such experiments. We also consider a `dirty' superconductor, in which the scattering with non-magnetic impurities is also much smaller than the superconducting coherence length.\\
\indent The Usadel equation comes from a quasiclassical approximation to the Gorkov equations. The Gorkov equations form a closed set of equations that describe the equations of motion for the Green's function of a device and incorporate all the results of microscopic BCS theory. This is done by introducing an anomalous Green's function to account for electron pairing in addition to the normal electronic Green's function. However, this is often difficult to work with, and the Usadel equation, which is derived from it is often used to describe 1D superconducting systems. Both the Gorkov equation, and by extension, the Usadel equation can easily account for dirty systems with impurities.\\
\indent We take into account spin-orbit and spin-flip scattering in the SC which may be intrinsic or arise from scattering off magnetic impurities. The MI layer also induces a uniform Zeeman field $V_Z^{SC}$ in the superconductor. In a dirty superconductor, the effect of the MI on the SC is microscopically equivalent to applying a magnetic field \cite{cottet2009spin}.
\newline
The Usadel equation in the Nambu spinor basis is: 
\begin{equation}
    D {\nabla} \cdot(\check{g} {\nabla} \check{g})-\left[\omega_{n} \hat{\tau}_{z}+i {V}_{Z}^{S C} \cdot \hat{\boldsymbol{\sigma}} \hat{\tau}_{z}+\Delta \hat{\tau}_{x}+\check{\Sigma}, \check{g}\right]=0,
\end{equation}
where, the quasi-classical Green's function for the bilayer is $\check{g}$. Here we have ignored the orbital effects, so the vector potential, i.e., $A=0$, $D$ is the diffusion constant, $V_Z^{SC}$ is the Zeeman field induced in the SC due to the MI including any additional external magnetic field, $\Delta$ is the superconducting gap, and $\check{\Sigma}$ represents the self energy due to scattering.\\
\indent The self energy due to scattering incorporates the spin-orbit ($\Sigma_{so}$) scattering and the spin-flip scattering ($\Sigma_{sf}$), which are characterised by the timescales $\tau_{so},\tau_{sf}$ respectively.
\begin{equation}
    \begin{split}
        \Sigma&=\Sigma_{s o}+\Sigma_{s f}\\
        \check{\Sigma}_{s o}&=\hat{\sigma} \check{g} \hat{\boldsymbol{\sigma}} /\left(8 \tau_{s o}\right)\\
        \check{\Sigma}_{s f}&=\hat{\boldsymbol{\sigma}} \hat{\tau}_{z} \check{g} \hat{\tau}_{z} \hat{\boldsymbol{\sigma}} /\left(8 \tau_{s f}\right).
    \end{split}
\end{equation}
We define the energy scale for the relaxation processes as:
$\Gamma_{so(sf)} = \frac{3}{2\tau_{so(sf)}}$.
Parameterizing the quasi-classical green's function in terms of $\theta$ and $\phi$, we can write:
\begin{equation}
     \check{g}\left(\omega_{n}, \boldsymbol{r}\right)=  \hat{\tau}_{z} \cos \theta\left(\cosh \phi+i \hat{\sigma}_{x} \tan \theta \sinh \phi\right)+\hat{\tau}_{x} \sin \theta\left(\cosh \phi-i \hat{\sigma}_{x} \cot \theta \sinh \phi\right).
\end{equation}
This yields the following equations:
\begin{equation}
 \begin{split} D \nabla^{2} \theta+2 \cosh \phi\left(\Delta \cos \theta-\omega_{n} \sin \theta\right)-2 V_{Z}^{S C} \sinh \phi \cos \theta-\frac{\Gamma_{s f}}{6}\left(2 \cosh ^{2} \phi+1\right) \sin 2 \theta &=0 \\-D \nabla^{2} \phi+2 \sinh \phi\left(\Delta \sin \theta+\omega_{n} \cos \theta\right)-2 V_{Z}^{S C} \cosh \phi \sin \theta+\left(\frac{2 \Gamma_{s o}}{3}+\frac{\Gamma_{s f}}{3} \cos 2 \theta\right) \cosh \phi \sinh \phi &=0 . \end{split}
\end{equation}
In our case, the effect of the MI and SM on the superconductor are uniform, and therefore $\check{g}(\omega,\textbf{r})$, and thus, $\theta(\omega,\textbf{r})$ and $\phi(\omega,\textbf{r})$ are only functions of $\omega$ and not $\textbf{r}$. This means that the diffusive term drops out of the Usadel equation, and we can simply solve the nonlinear algebraic equations for $\theta(\omega,\textbf{r})$ and $\phi(\omega,\textbf{r})$.
\begin{equation}
  \begin{split} 2 \cosh \phi\left(\Delta \cos \theta-\omega_{n} \sin \theta\right)-2 V_{Z}^{S C} \sinh \phi \cos \theta-\frac{\Gamma_{s f}}{6}\left(2 \cosh ^{2} \phi+1\right) \sin 2 \theta &=0 \\2 \sinh \phi\left(\Delta \sin \theta+\omega_{n} \cos \theta\right)-2 V_{Z}^{S C} \cosh \phi \sin \theta+\left(\frac{2 \Gamma_{s o}}{3}+\frac{\Gamma_{s f}}{3} \cos 2 \theta\right) \cosh \phi \sinh \phi &=0 . \end{split} 
\end{equation}

\begin{figure}
    \centering
    \includegraphics[scale = 0.75]{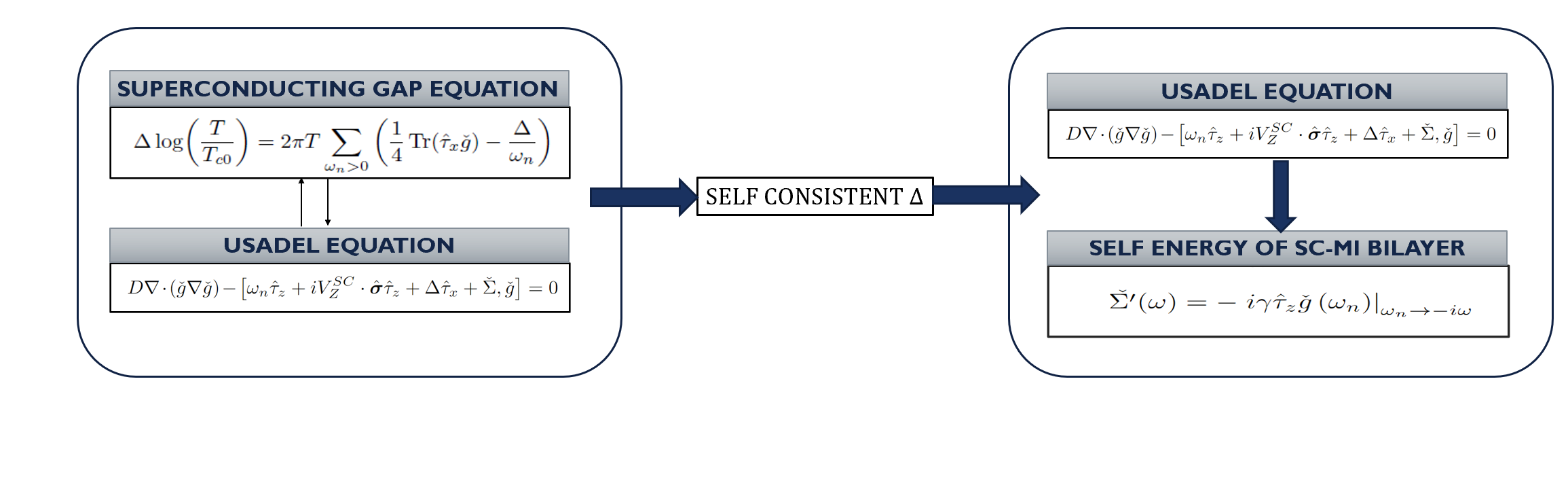}
    \caption{Flow of the calculation for the self energy $\Sigma'$ and the calculation of $\Delta$.The Usadel equation is solved with the superconducting gap equation and a thermodynamic constraint to yield the self-consistent value of $\Delta$. With this value of $\Delta$, the Usadel equation is solved again to obtain the quasi-classical Green's function for the bilayer $\check{g}$. The superconducting gap is induced in the bare Rashba nanowire by considering the proximity effect of the bilayer, which is taken into account by adding the $\Sigma^{'}$ term while finding the retarded Green's function of the nanowire.}
    \label{fig:schematic}
\end{figure}

\begin{equation}
\label{eq:final_sig}
    \check{\Sigma'}(\omega)=-\left.i \gamma \hat{\tau}_{z} \check{g}\left(\omega_{n}\right)\right|_{\omega_{n} \rightarrow-i \omega}.
\end{equation}
Since this bilayer is in contact with the entire Rashba nanowire, we consider the self energy to be $\Sigma'(\omega) = \mathbb{I}_{N}\otimes \check{\Sigma(\omega)'}$. \\
\indent For the Green's function calculations, we solve the Usadel equations in the real time domain. However, before we solve for the Green's function for the device, we use the Usadel equation to find the self-consistent value of the superconducting gap in the presence of external magnetic fields and the magnetic field due to the magnetic insulator.
\\
\indent
We solve the Usadel equation self-consistently with the superconducting gap equation \ref{eq:gap}, along with a thermodynamic constraint \ref{eq:thermodynamic} that the free energy of the superconducting phase should be greater than that of the normal metal phase, such that,
\begin{equation}
\label{eq:gap}
\Delta \log \left(\frac{T}{T_{c 0}}\right)=2 \pi T \sum_{\omega_{n}>0}\left(\frac{1}{4} \operatorname{Tr}\left(\hat{\tau}_{x} \check{g}\right)-\frac{\Delta}{\omega_{n}}\right),
\end{equation}
and 
\begin{equation}
\begin{split}
\label{eq:thermodynamic}
f_{s n}=\pi T \nu_{0} \sum_{\omega_{n}>0}\left\{4 \omega_{n}-2 \cosh \phi\left(2 \omega_{n} \cos \theta+\Delta \sin \theta\right)+4 V_{Z}^{S C} \sinh \phi \sin \theta+D\left[\nabla^{2} \theta-\nabla^{2} \phi\right]+\right. \\
\left.+\frac{1}{2}\left[\Gamma_{s o}+\Gamma_{s f}-\left(\Gamma_{s o}+\Gamma_{s f} \cos 2 \theta\right) \cosh ^{2} \phi-\frac{1}{3}\left(\Gamma_{s o}-\Gamma_{s f} \cos 2 \theta\right) \sinh ^{2} \phi\right]\right\}.
\end{split}
\end{equation}
We start with the initial guess of $\Delta$ being $\Delta_0$, calculate the Green's function for each $\omega_n$, find $\Delta$, and then repeat until it converges. The entire procedure for this calculation is schematized in Fig.~\ref{fig:schematic}.

\begin{figure}
    \centering
    \includegraphics[scale = 0.2]{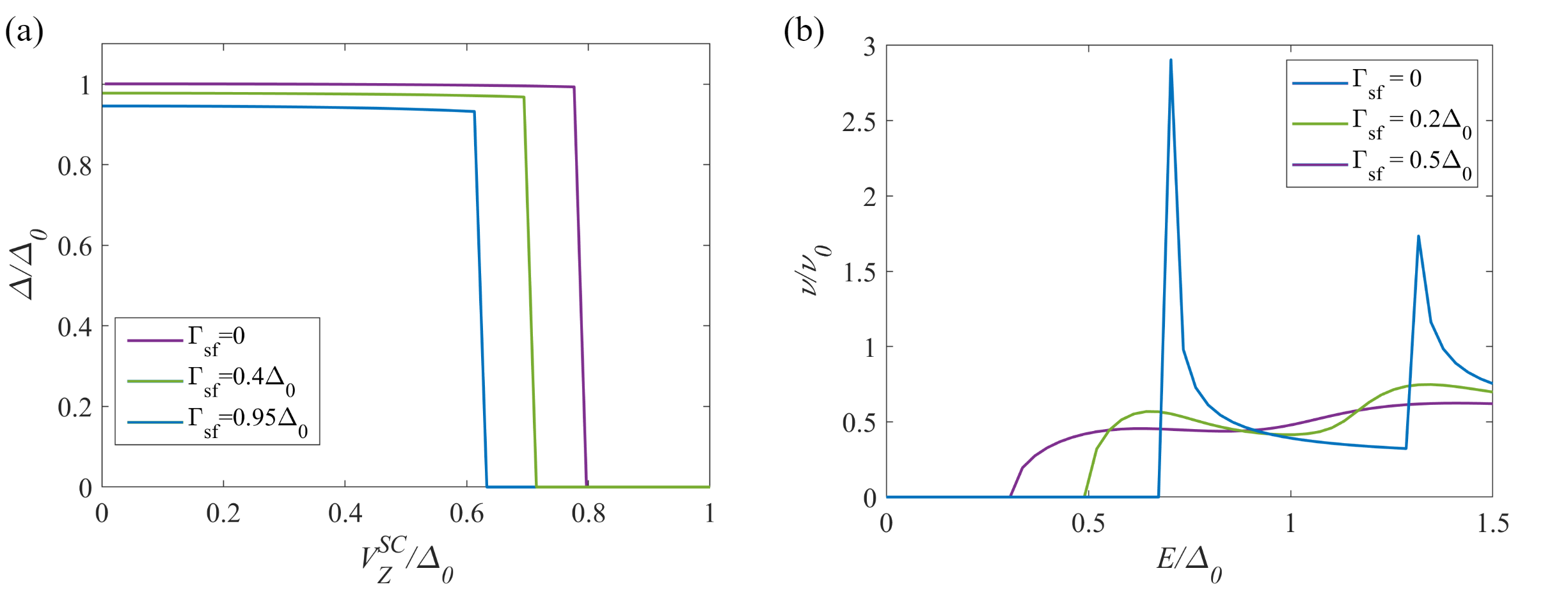}
    \caption{(a) Self consistent value of the superconducting gap $\Delta$ in terms of the bare superconducting gap $\Delta_{0}$ as the total magnetic field in the superconductor $V_{Z}^{SC}$ is varied. The critical field at which the superconducting gap goes to 0 (Clogston limit) decreases on increasing the spin flip scattering. (b) Density of states in the superconductor ($\nu$) is plotted as a function of energy, in terms of the normal density of states at the Fermi level ($\nu_{0}$). Here, we take $\Gamma_{so} = 0.4\Delta_{0}$ while varying $\Gamma_{sf}$}.
    \label{fig:Clogston}
\end{figure}

Figure~\ref{fig:Clogston} depicts the results of the self-consistent calculation for various values of the magnetic field and the scattering times. These results agree with \cite{khindanov2021topological}. The Chandrasekhar-Clogston limit places a stringent restriction on the maximum value of a magnetic field that can co-exist with superconductivity. The superconducting pairing potential undergoes a first-order transition into the normal state. This limit is clearly seen in our computational results. The Clogston limit increases with increase in spin-orbit coupling and decreases with decrease in spin-orbit coupling. 

\section*{Supplementary Note 2: The Keldysh NEGF approach}
\subsection{Current calculation}
\begin{figure}[h]
    \centering
    \includegraphics[scale = 0.75]{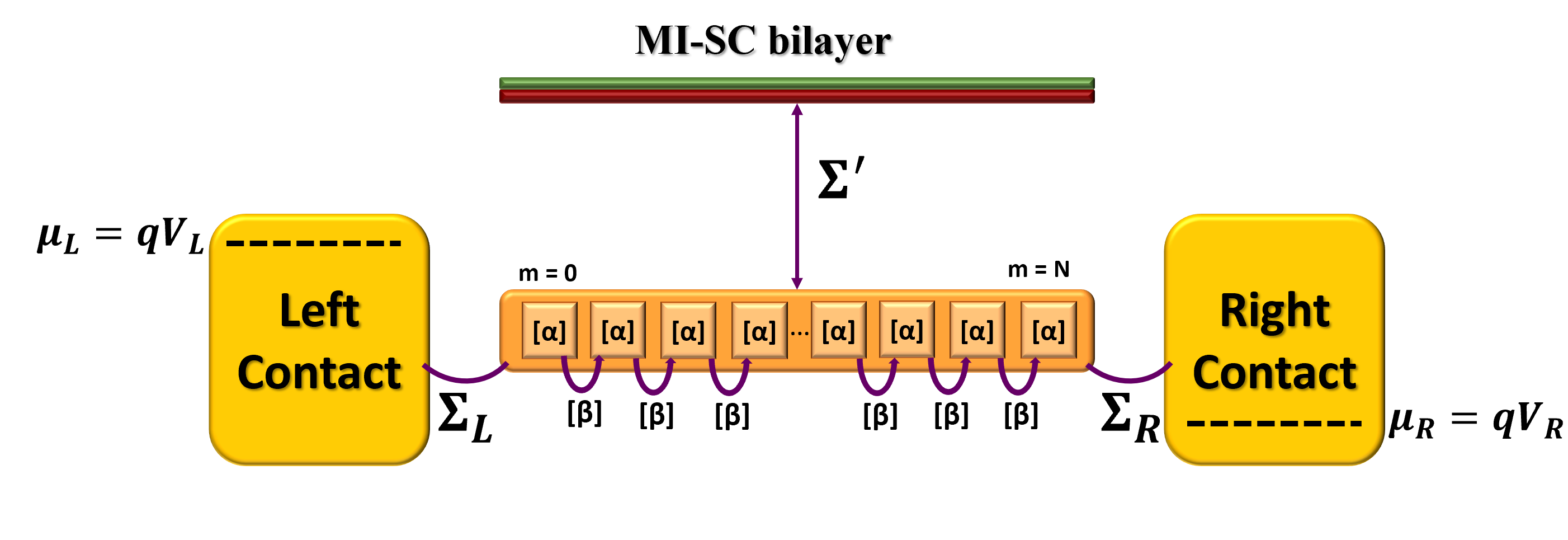}
    \caption{Schematic for the numerical calculations. The typical NEGF scheme involves the discretization of the channel Hamiltonian and expressing the coupling to various contacts using self-energies. In our setup, the voltages are applied at the left and right contacts, and the Usadel equations are employed to calculate the self energy due to the bilayer that is in contact with the Rashba nanowire channel. }
    \label{fig:my_label}
\end{figure}

 In the NEGF approach, the device is partitioned into a central channel, which comprises the Rashba nanowire and the leads as depicted in Fig.~\ref{fig:my_label}, which are incorporated using the self-energies. Here, the interface with the SC-MI bilayer, is also incorporated as a self-energy, and effectively acts as a third contact. This enables us to calculate the retarded Green's function for the channel as:
\begin{equation} \label{b1}
G^{r}(E)=\left[(E+i \eta) I-H_{SM}-\Sigma_{L}^{r}-\Sigma_{R}^{r} -\Sigma^{'}\right]^{-1},
\end{equation}
where $E$ is the free variable energy, and I is the identity
matrix of the dimension of the channel Hamiltonian, $\eta$ is a small positive damping parameter, and $\Sigma^{r}_{L}$ and $\Sigma^{r}_{R}$ represent the retarded self energies for the semi-infinite contacts and $\Sigma^{'}$ represents the self energy of the SC-MI bilayer, whose calculation was detailed in the previous section.

In the wide-band approximation, the self-energy of the contacts can be represented in their own eigenbasis, and are hence diagonal. They are characterised by a parameter $\gamma_{L/R}$, which represents the escape rate into the contacts. The current operator for an N-S-N device, derived from first principles in various references \cite{leumer2020linear,kejriwal2021can}, to discern the various components of the current:
\begin{equation} \label{b12}
\begin{aligned}
I_{L}^{e(h)}(E) &=\frac{e}{h}\left(\text { Trace }\left(\Gamma_{L}^{e e(h h)} G^{r} \Gamma_{R}^{e e(h h)} G^{a}\right)\left[f_{L}^{e e(h h)}(E)-f_{R}^{e e(h h)}(E)\right]\right) \longrightarrow (i)\\
&+\frac{e}{h}\left(\text { Trace }\left(\Gamma_{L}^{e e(h h)} G^{r} \Gamma_{L}^{h h(e e)} G^{a}\right)\left[f_{L}^{e e(h h)}(E)-f_{L}^{h h(e e)}(E)\right]\right) \longrightarrow (ii)\\
&+\frac{e}{h}\left(\text { Trace }\left(\Gamma_{L}^{e e(h h)} G^{r} \Gamma_{R}^{h h(e e)} G^{a}\right)\left[f_{L}^{e e(h h)}(E)-f_{R}^{h h(e e)}(E)\right]\right), \longrightarrow (iii)
\end{aligned}
\end{equation}
where the term (i) represents the direct transmission process of
either the electron or the hole, (ii) represents the direct
Andreev transmission and (iii) represents the crossed
Andreev transmission. At this point it is worth noting
that $f^{ee}_{\alpha} = f(E-\mu_{\alpha})$, $f^{hh}_{\alpha} = f(E+\mu_{\alpha})$, and that $\Gamma_{\alpha}$ is the imaginary part of the respective self energy. In our device, we need to consider the extra contribution to the current arising due to the presence of the effective third-contact \cite{kejriwal2021can}.
\subsection{Local and non-local conductance}
\label{appendix:C}
\subsubsection{Floating superconductor configuration}
For a floating SC, we consider the only effect of the SC is proximity induced superconductivty, and we do not consider as a terminal for current flow. We derive the expression for the current in a generic bias situation, ($V_{L}$,$V_{R}$) for the left and right contacts with $\mu_{L}=eV_{L}$ and $\mu_{R}=eV_{R}$
\\
\indent
Both the electron and hole flows contribute to the current at any contact. The net current at the left contact is given by:
\begin{equation}
I^{L}=\frac{I_{L}^{(e)}-I_{L}^{(h)}}{2}.
\end{equation}
The individual electron and hole components can then be derived using:
\begin{equation}
\begin{aligned}
I_{L}^{(e)}=\frac{e}{h} &\left\{\int d E T_{A}^{(e)}(E)\left[f\left(E-e V_{L}\right)-f\left(E+e V_{L}\right)\right]\right.\\
&+\int d E T_{C A R}^{(e)}(E)\left[f\left(E-e V_{L}\right)-f\left(E+e V_{R}\right)\right] \\
&\left.+\int d E  T_{D}^{(e)}(E)\left[f\left(E-e V_{L}\right)-f\left(E-e V_{R}\right)\right]\right\},
\end{aligned}
\end{equation}

\begin{equation}
\begin{aligned}
I_{L}^{(h)}=\frac{e}{h} &\left\{\int d E T_{A}^{(h)}(E)\left[f\left(E+e V_{L}\right)-f\left(E-e V_{L}\right)\right]\right.\\
&+\int d E T_{C A R}^{(h)}(E)\left[f\left(E+e V_{L}\right)-f\left(E-e V_{R}\right)\right] \\
&\left.+\int d E T_{D}^{(h)}(E)\left[f\left(E+e V_{L}\right)-f\left(E+e V_{R}\right)\right]\right\}.
\end{aligned}
\end{equation}
Here, $I^{(e)}_L$ and $I^{(h)}_L$ can be shown to be equal and opposite using symmetry conditions and especially:
\begin{equation}
\begin{aligned}
T_{CAR}^{(e)}(E) \equiv T^{(h)}_{C A R}(E), \mspace{50mu}
T_{D}^{(e)}(E) \equiv T_{D}^{(h)}(-E).
\end{aligned}
\end{equation}
Therefore, it suffices to calculate either one of $I^{(e)}_L$ and $I^{(h)}_L$ in order to calculate the net current from a contact. We also consider the conductance matrix, which explicitly evaluates the local and non-local components of the conductance. 
\begin{equation}
\mathrm{G}=\left(\begin{array}{cc}
G_{L L} & G_{L R} \\
G_{R L} & G_{R R}
\end{array}\right)=\left(\begin{array}{cc}
\left.\frac{\partial I_{L}}{\partial V_{L}}\right|_{V_{R}=0} & \left.\frac{\partial I_{L}}{\partial V_{R}}\right|_{V_{L}=0} \\
\left.\frac{\partial I_{R}}{\partial V_{L}}\right|_{V_{R}=0} & \left.\frac{\partial I_{R}}{\partial V_{R}}\right|_{V_{L}=0}.
\end{array}\right)
\end{equation}
Henceforth, we consider only the electron current since $I_L=I^{(e)}_L$. The local conductance is then given as:
\begin{equation}
\begin{aligned}
G_{L L} &\equiv \left. \frac{\partial I_{L}}{\partial V_{L}} \right|_{V_{R}\equiv 0 }\\
& \equiv \frac { e } { h } \left\{\frac { \partial } { \partial V_L } \left[\int d E T_{A}(E)[f(E-e V_L)-f(E+eV_L)]\right.\right.
+\int d E T_{CAR}(E)[f(E-e V_L)-f(E)] \\
&\mspace{450mu}+\left.\left.\int d E T_{D}(E)[f(E-eV_L)-f(E)]\right]\right\} \\
& \equiv \frac{e}{h}\left[\int d E\right. T_{A}(E)\left[\frac{\partial f\left(E-eV_L\right)}{\partial V_L}-\frac{\partial f(E+e V_L)}{\partial V_L}\right] 
+ \int d E T_{CA R}(E)\left[\frac{\partial f(E-e V_L)}{\partial V_L}\right] \\
&\mspace{450mu} +\left.\int d E T_{D}(E)\left[\frac{\partial f(E-e V_L)}{\partial V_L}\right]\right].
\end{aligned}
\end{equation}
At zero temperature, the Fermi distribution function, $f\left(x\right)=\Theta\left(-x\right)$ where $\Theta$ is the Heaviside step function. This implies $f\left(E-eV \right)=\Theta\left(eV-E\right)$. Utilising the fact that the derivative of the Heaviside function is the Dirac delta function, we get:
\begin{equation}
\begin{aligned} \label{b22}
\left.G_{LL}(V)\right|_{T \rightarrow 0} \equiv \frac{e^2}{h}\left[T_{A}(E=e V)+T_{A}(E=-e V)\right.+
\left.T_{C A R}(E=eV)+T_{D}(E=e V)\right].
\end{aligned}
\end{equation}
The expression for non-local conductance can be derived similarly. 
\begin{equation}
\begin{aligned}
G_{LR}=& \left.\frac{\partial I_L}{\partial V_R}\right|_{V_L\equiv 0} \\
=&\frac { e } { h } \left\{\frac { \partial } { \partial V_R } \left[\int d E T_{CAR}(E)[f(E)-f(E+eV_R)]\right.\right.
\left.\left.+\int d  E T_{D}(E)\left[f\left(E\right)-f\left(E-e V_{R}\right)\right]\right]\right\} \\
=& \frac{e}{h}\left[ \int d E T_{CA R}(E)\left[\frac{-\partial f(E+e V_R)}{\partial V_R}\right]\right. 
+\left. \int d E T_{ D}(E)\left[\frac{-\partial f(E-e V_R)}{\partial V_R}\right].
\right]
\end{aligned}
\end{equation}
Once again, using the property of Fermi Dirac functions as $T\rightarrow 0$ as $f(x)=\Theta(-x)$ we get:
\begin{equation}
\begin{aligned} \label{b24}
\left.G_{LR}(V)\right|_{T \rightarrow 0} \equiv \frac{e^2}{h}\left[T_{CAR}(E=-e V)-T_{D}(E=e V)\right].
\end{aligned}
\end{equation}
\subsubsection{Grounded superconductor configuration}
For our device setup, current can flow into the SC-MI bilayer terminal as well. This leads to an additional self-energy term ($\Sigma^{'}$) in the Green's function. The Green's function for the grounded setup is thus given by :

\begin{equation}
G^{r}(E)=\left[(E+i \eta) I-H_{SM}-\Sigma_{L}^{r}-\Sigma_{R}^{r}-\Sigma^{'}\right]^{-1},
\end{equation}
where $\Sigma^{'}$ can be derived as outlined in \ref{eq:final_sig}.
where $\gamma$ represents the coupling strength between the bilayer and the Rashba nanowire.

In the presence of the superconducting contact, an additional term gets added to the previously derived formula for local conductance (\ref{b22}), and no additional term gets added to the non-local conductance formula (\ref{b24}).

We once again start with the general current operator for the left contact, and obtain the electron current for the left contact as:

\begin{equation} \label{b27}
\begin{aligned}
I_{L}^{e}(E)&=\frac{e}{h} \operatorname{Tr}\left[\Sigma_{L}^{<}(E)\left(G^{r}(E)-G^{a}(E)\right)\right.
\left.- G^{<}(E)\left(\Sigma_{L}^{a}-\Sigma_{L}^{r}\right)\right] \\
&=\frac{e}{h}  \operatorname{Tr}\left[\Gamma_{L} f_{L}\left(G^{r}(E)\Gamma G^{a}(E)\right)\right.
\left.+ i G^{r}(E)\left(\Sigma_{L}^{<}+\Sigma_{R}^{<}+\Sigma^{'<}\right) G^{a}(E) \Gamma_{L}\right],
\end{aligned}
\end{equation}
The broadening matrix $\Gamma$, represents the total broadening and is the sum of the broadening matrices for left,right and the SC-MI bilayer contact: $\Gamma = \Gamma_L+\Gamma_R+\Gamma^{'}$. On further expanding the terms in (\ref{b27}) we get,
\begin{equation}
I_{L}^{e}(E)=\frac{e}{h}\operatorname{Tr}\left[\Gamma_{L} f_{L}G^{r}(E)\left(\Gamma_L+\Gamma_R+\Gamma^{'}\right)G^{a}(E)+ iG^{r}(E)\left(i\Gamma_{L}f_{L}(E)+i\Gamma_{R}f_{R}(E)+i\Gamma^{'}f(E)\right)G^{a}(E)\Gamma_{L}\right].
\end{equation}
The additional terms due to the presence of the superconducting contact are:
\begin{equation} \label{b29}
\frac{e}{h}\operatorname{Tr}\left[\underbrace{\Gamma_{L} f_{L}G^{r}(E)\Gamma_SG^{a}(E)}_{(\mathrm{i})}-\underbrace{G^{r}(E)\Gamma^{'} f_{S}(E)G^{a}(E)\Gamma_{L}}_{(\mathrm{ii})} \right].
\end{equation}
The term (i) here has a $V_L$ dependence since $f_L=f(E-eV_L)$, whereas the term (ii) has neither $V_L$ nor $V_R$ dependence. Since the first term has a $V_L$ dependence it adds an additional term to the local conductance formula (\ref{b22}):
\begin{equation}
G^{'}_{LL}(V_L) = \frac{e}{h}\operatorname{Tr}\left[\Gamma_{L} \frac{df_{L}}{dV_L}G^{r}(E)\Gamma^{'}G^{a}(E)\right].
\end{equation}
Just as in the previous section, we utilize the fact that the derivative of Fermi function at zero temperature is the Dirac-delta function to get an extra term in the local conductance to:
\begin{equation}
G^{'}_{LL}(V) = \frac{e}{h}\operatorname{Tr}\left[\Gamma_{L}G^{r}(V)\Gamma^{'}G^{a}(V)\right].
\end{equation}
Since neither of the terms in (\ref{b29}) have $V_R$ dependence, no additional term gets added to the non-local conductance equation (\ref{b24}) on introducing the extra terminal.
\twocolumngrid
\end{document}